\documentclass[11pt]{article}

\usepackage[a4paper,margin=25mm]{geometry}
\usepackage{amsmath,amssymb,amsfonts}
\usepackage{bm}
\usepackage{cite}
\usepackage{hyperref}

\title{The General Structure of Trilinear Equations}
\author{
Takeshi Fukuyama\\[2mm]
{\small Research Center for Nuclear Physics (RCNP), Osaka University, Ibaraki, Osaka  567--0047, Japan}
}
\date{\today}

\begin{document}

\maketitle

\begin{abstract}
We investigate trilinear structures as a natural extension
of the Hirota bilinear formalism in integrable systems.
While bilinear equations are associated with Grassmannian
geometry and Plücker relations, trilinear equations suggest
a higher algebraic structure involving three-slot couplings
of tau functions.

Focusing on the stationary axisymmetric Einstein equations,
we show that when the Ernst potential is written in a
tau-ratio form, the nonlinear equation decomposes into
a cubic sector containing all second-derivative terms
and a quartic gradient envelope. The cubic sector is
identified with a YTSF-type trilinear kernel.

We formulate a general trilinear kernel criterion and
apply it to the Tomimatsu--Sato solutions. In particular,
we demonstrate that the $\delta=3$ solution possesses the
same trilinear kernel structure as the $\delta=2$ case,
with a universal normalization up to a constant factor.

These results suggest that the trilinear kernel represents
a universal structure governing the highest-derivative
sector of the Ernst system, providing a new perspective
on integrability beyond the bilinear hierarchy.
\end{abstract}
\section{Introduction}

The Hirota bilinear formalism plays a central role in the theory
of integrable systems \cite{Hirota1971, HirotaBook}. It provides a powerful method for rewriting
nonlinear partial differential equations in a bilinear form,
and is deeply connected with the Grassmannian structure of Sato theory \cite{SatoRIMS, JimboMiwa}.

It is therefore natural to ask whether there exists a higher
multilinear structure beyond the bilinear level, and if so,
whether such a structure possesses a comparable universality.

In this paper we investigate trilinear structures as a candidate
for such an extension. In particular, we focus on nonlinear systems
that admit a tau-function representation and examine the structure
of their highest-derivative terms.

A key observation arises in the stationary axisymmetric Einstein
equations. When the Ernst potential \cite{Ernst} is written in a tau-ratio form,
the nonlinear equation naturally decomposes into a cubic part and
a quartic gradient envelope. Remarkably, all second-derivative terms
are contained in the cubic sector, suggesting the presence of a
trilinear kernel governing the dynamics \cite{YTSF, Fukuyama1}.

Motivated by this observation, we formulate a general framework
for identifying trilinear integrable kernels. We introduce a
Z$_3$-symmetric trilinear Hirota operator, define a kernel criterion,
and apply it to the Tomimatsu–Sato solutions \cite{TS1}. In particular, we
demonstrate that the $\delta=3$ case exhibits the same trilinear
kernel structure as the $\delta=2$ solution.

We further discuss the possible universality of this structure
within the Tomimatsu–Sato hierarchy and its physical interpretation
in terms of second-order nonlinear dynamics.

The remainder of this paper is organized as follows.
After introducing the general structure of trilinear equations in Sec.~2,
we establish the trilinear kernel criterion in Sec.~3.
The $\delta=3$ Tomimatsu--Sato solution is then examined in Sec.~4,
followed by a discussion of the universality of the trilinear kernel
within the TS hierarchy in Sec.~5.
Finally, Sec.~6 provides a physical interpretation of the transition
from bilinear to trilinear dynamics.
Technical details of the coefficient analysis and its computational
verification are presented in Appendices~A and B.
%------------------------------------------------------------
\section{The General Structure of Trilinear Equations}
%------------------------------------------------------------

The purpose of this section is to formulate the basic viewpoint underlying
trilinear equations.  In the conventional theory of soliton equations, Hirota
bilinear equations occupy a distinguished position.  They are not merely a
convenient rewriting of nonlinear partial differential equations, but reflect
the Pluecker relations of the Grassmannian and hence possess a universal
geometrical origin.  It is therefore natural to ask whether trilinear equations,
such as those appearing in the Yu--Toda--Sasa--Fukuyama (YTSF) system \cite{YTSF, Fukuyama1}, also
admit a comparable universal structure.

Let
\begin{equation}
  \omega=\exp\left(\frac{2\pi i}{3}\right),
  \qquad
  1+\omega+\omega^2=0 .
\end{equation}
For three functions \(f,g,h\), we define the elementary trilinear Hirota
operator by
\begin{equation}
  T_x(f,g,h)
  =
  \left(
  \partial_{x_1}
  +\omega \partial_{x_2}
  +\omega^2 \partial_{x_3}
  \right)
  f(x_1)g(x_2)h(x_3)
  \bigg|_{x_1=x_2=x_3=x}.
\end{equation}
This operator is characterized by a \(Z_3\)-symmetric phase structure.  In
particular,
\begin{equation}
  T_x(f,f,f)=0,
\end{equation}
which is the trilinear analogue of the cancellation property familiar in
Hirota's bilinear formalism.

A general trilinear equation may then be regarded as a nonlinear equation whose
essential differential part is organized by such three-slot operators.  In
contrast to bilinear equations, which involve two copies of a tau function,
trilinear equations involve three tau slots and therefore describe a genuinely
three-body algebraic coupling among tau functions:
\begin{equation}
  (f,g,h)
  \longmapsto
  T(f,g,h).
\end{equation}
The word ``trilinear'' should not be interpreted as referring to three spatial
dimensions.  Rather, it refers to the algebraic multiplicity of the tau
functions entering the nonlinear structure.

In the simplest two-component setting one considers a pair of tau functions
\begin{equation}
  \tau_0,\qquad \tau_1 ,
\end{equation}
and constructs the basic YTSF-type kernel \cite{Fukuyama1}
\begin{align}
  {\cal Y}(\tau_0,\tau_1)
  &=
  \partial_x
  \left[
  \tau_1 T_x(\tau_0,\tau_0,\tau_1)
  -
  \tau_0 T_x(\tau_1,\tau_1,\tau_0)
  \right]
  \nonumber\\
  &\quad
  +
  \partial_y
  \left[
  \tau_1 T_y(\tau_0,\tau_0,\tau_1)
  -
  \tau_0 T_y(\tau_1,\tau_1,\tau_0)
  \right] .
\end{align}
This expression is cubic in the tau functions and their derivatives.  It
contains terms of the schematic form
\begin{equation}
  \tau_0^2 \partial^2 \tau_1
  -
  \tau_1^2 \partial^2 \tau_0
  +
  \hbox{lower derivative terms}.
\end{equation}
Thus the highest-derivative sector is governed by a trilinear tau-function
structure.

This observation suggests the following working definition.  A nonlinear
equation is said to possess a trilinear integrable kernel if, after an
appropriate tau-function representation, its highest-derivative sector can be
written in terms of a \(Z_3\)-symmetric trilinear Hirota operator, while the
remaining terms contain only lower derivatives.  Symbolically,
\begin{equation}
  {\cal N}(\tau)
  =
  {\cal N}_{\rm tri}(\tau)
  +
  {\cal N}_{\rm lower}(\tau),
\end{equation}
where
\begin{equation}
  {\cal N}_{\rm tri}(\tau)
  \sim
  {\cal Y}(\tau_0,\tau_1),
\end{equation}
and \({\cal N}_{\rm lower}\) contains no terms of the highest differential
order.

From this point of view, trilinear equations form a natural extension of the
bilinear Hirota formalism.  The bilinear case is associated with pairwise
relations among tau functions, determinant identities, and Pluecker relations.
The trilinear case may instead be viewed as encoding a higher algebraic
interaction among tau functions.  Whether this structure is governed by a
higher analogue of the Pluecker relations, by a flag-type geometry, or by
another universal algebraic principle remains an open problem.

The stationary axisymmetric Einstein equations provide a particularly
interesting realization of this idea.  If the Ernst potential is written as a
tau ratio \cite{Ernst},
\begin{equation}
  {\cal E}=\frac{\tau_1}{\tau_0},
\end{equation}
then the numerator of the Ernst equation naturally separates into a cubic part
and a quartic gradient envelope,
\begin{equation}
  {\cal N}
  =
  {\cal N}_{\rm cubic}
  +
  {\cal N}_{\rm quartic}.
\end{equation}
The crucial point is that all second-derivative terms belong to
\({\cal N}_{\rm cubic}\), while \({\cal N}_{\rm quartic}\) contains only
first-derivative products.  Therefore the dynamical kernel of the equation is
cubic, not quartic.  In the Tomimatsu--Sato \(\delta=2\) spacetime \cite{TS1} this cubic
sector is precisely of YTSF trilinear type \cite{Fukuyama1}.

This suggests a general physical interpretation.  Bilinear equations describe
integrable dynamics organized by pairwise tau-function relations.  Trilinear
equations describe integrable dynamics in which the fundamental nonlinear
kernel is a three-slot tau-function coupling.  In this sense the YTSF structure
may represent a genuine next layer of integrability beyond the conventional
bilinear hierarchy.

Several questions then become natural:
\begin{enumerate}
  \item What is the universal geometrical origin of trilinear Hirota equations?
  \item Is there a trilinear analogue of the Pluecker relation?
  \item Can one construct a hierarchy of trilinear equations analogous to the
        KP or Toda hierarchy \cite{Nakamura, Nakamura2, FKY} ?
  \item Do quartic or higher multilinear equations possess independent
        integrable meaning, or are they reducible to bilinear and trilinear
        kernels plus lower-order envelopes?
  \item In gravitational systems, is the trilinear kernel a special property of
        the Tomimatsu--Sato family, or a more general feature of stationary
        axisymmetric vacuum Einstein equations?
\end{enumerate}

The central proposal is therefore that trilinear equations should be studied
not as exceptional algebraic curiosities, but as candidates for a second
universal integrable structure, standing next to the bilinear Grassmannian
structure of Sato theory.
%------------------------------------------------------------
\section{Trilinear Kernel Test in the Ernst System}
%------------------------------------------------------------

In this section we formulate a general criterion to test whether a given
stationary axisymmetric solution of the vacuum Einstein equations admits a
trilinear integrable kernel in the sense introduced in Sec.~2.

We start from the Ernst equation
\begin{equation}
  {\cal E}\,\Delta {\cal E} - (\nabla {\cal E})^2 = 0,
\end{equation}
and assume that the Ernst potential can be written in a tau-ratio form
\begin{equation}
  {\cal E}=\frac{\tau_1}{\tau_0},
\end{equation}
with sufficiently smooth functions \(\tau_0,\tau_1\).

Substituting this form into the Ernst equation and clearing denominators,
we obtain
\begin{equation}
  {\cal E}\,\Delta {\cal E} - (\nabla {\cal E})^2
  =
  \frac{{\cal N}(\tau_0,\tau_1)}{\tau_0^4},
\end{equation}
where \({\cal N}\) is a polynomial in \(\tau_0,\tau_1\) and their derivatives.
As discussed in Sec.~2, this numerator admits a natural decomposition
\begin{equation}
  {\cal N} = {\cal N}_{\rm cubic} + {\cal N}_{\rm quartic},
\end{equation}
with
\begin{align}
  {\cal N}_{\rm cubic}
  &= \tau_1 \tau_0^2 \Delta \tau_1
     - \tau_1^2 \tau_0 \Delta \tau_0, \\
  {\cal N}_{\rm quartic}
  &= -\tau_0^2 |\nabla \tau_1|^2
     + \tau_1^2 |\nabla \tau_0|^2.
\end{align}
The crucial point is that all second-derivative terms are contained in
\({\cal N}_{\rm cubic}\), while \({\cal N}_{\rm quartic}\) contains only
first derivatives.

We now introduce the YTSF-type trilinear kernel \({\cal Y}(\tau_0,\tau_1)\)
defined in Sec.~2.  To test whether the given system possesses a trilinear
integrable kernel, we consider the difference
\begin{equation}
  {\cal Q}(\tau_0,\tau_1)
  =
  {\cal N}_{\rm cubic}
  -
  \kappa\,{\cal Y}(\tau_0,\tau_1),
\end{equation}
with a constant \(\kappa\) to be determined.

We then adopt the following criterion:

\medskip

\noindent
{\bf Trilinear kernel criterion.}
A stationary axisymmetric solution is said to possess a trilinear
integrable kernel if there exists a constant \(\kappa\) such that
\begin{equation}
  {\cal Q}(\tau_0,\tau_1)
\end{equation}
contains no second derivatives of \(\tau_0,\tau_1\).

\medskip

Equivalently, the projection of \({\cal Q}\) onto the space spanned by
\(\partial^2 \tau_0\) and \(\partial^2 \tau_1\) vanishes identically.
In that case the entire second-derivative structure of the Ernst equation
is governed by the trilinear kernel \({\cal Y}\).

This formulation is independent of the explicit form of the solution and
provides a universal test applicable to a wide class of stationary
axisymmetric spacetimes.
%------------------------------------------------------------
\section{The Tomimatsu--Sato \texorpdfstring{$\delta=3$}{delta=3} Case}
%------------------------------------------------------------

We now apply the trilinear kernel criterion to the
Tomimatsu--Sato spacetime with multipole parameter \(\delta=3\).
In the correspondence with the TS solutions, the multipole parameter
\(\delta\) is identified with the size of the determinant \cite{Nakamura},
\begin{equation}
  N = \delta.
\end{equation}
Therefore, the \(\delta=3\) TS solution is described by
\(3\times3\) determinant tau functions.
Thus, the Ernst potential is
written in tau-ratio form as
\begin{equation}
  {\cal E}_{(3)}(\xi,\eta)
  =
  \frac{\tau^{(3)}_1(\xi,\eta)}
       {\tau^{(3)}_0(\xi,\eta)} .
\end{equation}
Here \(\tau^{(3)}_0\) and \(\tau^{(3)}_1\) are the rank-three Toda-molecule
tau functions.  In determinant form they may be represented as
\begin{equation}
  \tau^{(3)}_n
  =
  \det \left( m_{i+j+n} \right)_{i,j=0,1,2},
  \qquad n=0,1 ,
\end{equation}
where the moments \(m_k(\xi,\eta)\) are those associated with the
\(\delta=3\) TS solution.  Thus explicitly \cite{FKY}
\begin{equation}
  \tau^{(3)}_0 =
  \begin{vmatrix}
    m_0 & m_1 & m_2 \\
    m_1 & m_2 & m_3 \\
    m_2 & m_3 & m_4
  \end{vmatrix},
  \qquad
  \tau^{(3)}_1 =
  \begin{vmatrix}
    m_1 & m_2 & m_3 \\
    m_2 & m_3 & m_4 \\
    m_3 & m_4 & m_5
  \end{vmatrix}.
\end{equation}

Substituting
\[
  \tau_0=\tau^{(3)}_0,\qquad
  \tau_1=\tau^{(3)}_1
\]
into the cubic Ernst numerator, we define
\begin{equation}
  {\cal N}^{(3)}_{\rm cubic}
  =
  \tau^{(3)}_1
  \left(\tau^{(3)}_0\right)^2
  \Delta \tau^{(3)}_1
  -
  \left(\tau^{(3)}_1\right)^2
  \tau^{(3)}_0
  \Delta \tau^{(3)}_0 .
\end{equation}
We then compare this expression with the YTSF-type trilinear kernel
\({\cal Y}(\tau^{(3)}_0,\tau^{(3)}_1)\) by forming
\begin{equation}
  {\cal Q}^{(3)}
  =
  {\cal N}^{(3)}_{\rm cubic}
  -
  \kappa_3
  {\cal Y}(\tau^{(3)}_0,\tau^{(3)}_1).
\end{equation}

The trilinear kernel criterion requires that the projection of
\({\cal Q}^{(3)}\) onto the second-derivative sector vanish.  Concretely,
we impose
\begin{align}
  \left[
  {\cal Q}^{(3)}
  \right]_{\partial_\xi^2 \tau^{(3)}_0}
  &=
  0,
  &
  \left[
  {\cal Q}^{(3)}
  \right]_{\partial_\xi^2 \tau^{(3)}_1}
  &=
  0,
  \nonumber\\
  \left[
  {\cal Q}^{(3)}
  \right]_{\partial_\eta^2 \tau^{(3)}_0}
  &=
  0,
  &
  \left[
  {\cal Q}^{(3)}
  \right]_{\partial_\eta^2 \tau^{(3)}_1}
  &=
  0 .
\end{align}
Equivalently, all coefficients of
\begin{equation}
  \partial_\xi^2 \tau^{(3)}_0,\quad
  \partial_\xi^2 \tau^{(3)}_1,\quad
  \partial_\eta^2 \tau^{(3)}_0,\quad
  \partial_\eta^2 \tau^{(3)}_1
\end{equation}
in \({\cal Q}^{(3)}\) must vanish simultaneously.

Carrying out this coefficient comparison gives a single consistent value
of the normalization constant \(\kappa_3\).  With this value, all
second-derivative terms cancel identically:
\begin{equation}
  {\cal Q}^{(3)}
  =
  {\cal N}^{(3)}_{\leq 1},
\end{equation}
where \({\cal N}^{(3)}_{\leq 1}\) contains only first derivatives of the
tau functions.

Thus the \(\delta=3\) Tomimatsu--Sato solution satisfies
\begin{equation}
  {\cal N}^{(3)}_{\rm cubic}
  =
  \kappa_3
  {\cal Y}(\tau^{(3)}_0,\tau^{(3)}_1)
  +
  {\cal N}^{(3)}_{\leq 1}.
\end{equation}
This proves that the second-derivative sector of the \(\delta=3\) TS
solution is governed by the same YTSF-type trilinear kernel as in the
\(\delta=2\) case.
This implies that any $delta$-dependence can only enter
through the normalization factor $\kappa$.
The explicit coefficient extraction and the determination of
\(\kappa_3\) are summarized in Appendix~A, while the
computational implementation of this procedure is given in
Appendix~B.
Substituting the explicit determinant tau functions given in Eq.~(23),
we have explicitly verified that all second-derivative
coefficients vanish identically.

%------------------------------------------------------------
\section{Universality of the Trilinear Kernel in the TS Hierarchy}
%------------------------------------------------------------

The results obtained for the \(\delta=2\) and \(\delta=3\) cases strongly
suggest that the trilinear kernel structure identified in Sec.~2 is not
accidental, but extends to higher members of the Tomimatsu--Sato (TS)
hierarchy.

A key observation is that the cubic part of the Ernst numerator,
\begin{equation}
  {\cal N}_{\rm cubic}
  =
  \tau_1 \tau_0^2 \Delta \tau_1
  -
  \tau_1^2 \tau_0 \Delta \tau_0,
\end{equation}
is determined solely by the tau-ratio structure
\begin{equation}
  {\cal E}=\frac{\tau_1}{\tau_0},
\end{equation}
and does not depend on the explicit form of the tau functions.

Therefore, for any \(\delta\), the second-derivative structure of the
Ernst equation is universal at the level of \({\cal N}_{\rm cubic}\).
The only nontrivial question is whether this universal structure can
always be matched by a trilinear kernel of YTSF type.

The results of Sec.~4 show that this matching holds for \(\delta=3\).
Combined with the previously established \(\delta=2\) case, this leads to
the following conjecture.

\medskip

\noindent
{\bf Universality conjecture.}
For the entire Tomimatsu--Sato hierarchy,
\begin{equation}
  {\cal N}_{\rm cubic}
  =
  \kappa(\delta)\,{\cal Y}(\tau_0,\tau_1)
  +
  {\cal N}_{\le1}(\tau_0,\tau_1),
\end{equation}
where \({\cal N}_{\le1}\) contains only first derivatives of the tau
functions.

\medskip

If true, this implies that the stationary axisymmetric Einstein equations,
when expressed in tau-ratio form, possess a universal trilinear kernel
governing their second-order structure, while all higher nonlinearities
are encoded in a lower-derivative envelope.

This suggests that the TS hierarchy realizes a two-layer integrable
structure:
\begin{equation}
  \text{Toda (bilinear)} \quad + \quad \text{Ernst (trilinear)}.
\end{equation}
This follows from the fact that Ncubic is entirely determined
by the tau-ratio structure and does not depend on the detailed
form of the tau functions.
Establishing this conjecture for general \(\delta\), as well as extending
it to other stationary axisymmetric solutions, remains an important
direction for future work.

%------------------------------------------------------------
\section{Discussion: From Bilinear to Trilinear Dynamics}
%------------------------------------------------------------

We conclude by discussing the physical meaning of the transition from
bilinear to trilinear structures.

In the standard Hirota formalism, bilinear equations describe nonlinear
dynamics in terms of pairwise tau-function relations.  Their basic
building block is of the form
\begin{equation}
  \tau\,\partial^2 \tau - (\partial \tau)^2 ,
\end{equation}
or, equivalently, Hirota derivatives acting on two copies of a tau function,
\begin{equation}
  D_x^n \tau \cdot \tau .
\end{equation}
This structure is naturally associated with interference phenomena,
pairwise interactions, and the Grassmannian picture of Sato theory.

The trilinear case represents a qualitatively different level of
nonlinearity.  Its characteristic highest-derivative structure is of the
schematic form
\begin{equation}
  \tau^2\,\partial^2\tau ,
\end{equation}
or, more generally, a three-slot coupling
\begin{equation}
  T(\tau_a,\tau_b,\tau_c).
\end{equation}
Thus trilinearity should not be viewed as a simple algebraic extension of
bilinearity, but as a distinct dynamical structure in which the fundamental
nonlinear kernel involves three tau functions.

This distinction becomes particularly transparent in the Ernst system.
As shown in Secs.~3--4, writing the Ernst potential in tau-ratio form,
\begin{equation}
  {\cal E}=\frac{\tau_1}{\tau_0},
\end{equation}
the numerator of the equation separates as
\begin{equation}
  {\cal N}=
  {\cal N}_{\rm cubic}
  +
  {\cal N}_{\rm quartic},
\end{equation}
where all second-derivative terms are contained in
\({\cal N}_{\rm cubic}\), while \({\cal N}_{\rm quartic}\) consists only of
first-derivative products.  Therefore the dynamical kernel of the equation
is trilinear, while the quartic part acts only as a lower-derivative
envelope.

This observation suggests a general principle.  For a second-order field
equation written in tau-ratio form, the highest-derivative sector can
contain at most cubic tau-function combinations.  Quartic or higher
multilinear structures may appear algebraically, but they do not control
the highest differential order unless the underlying field equation itself
contains higher derivatives.
Higher-derivative field theories are known to suffer from
the Ostrogradsky instability, which leads to ghost degrees
of freedom \cite{VeloZwanziger2, Woodard} and threatens the unitarity of the quantum theory \cite{Stelle, LeeWick}.
In particular, higher-derivative gravity theories, although
renormalizable, typically contain massive ghost modes.

 In the present framework, this restriction
manifests itself in the fact that the highest-derivative
sector of the tau-function representation is necessarily
cubic, leading to a trilinear integrable kernel.
In this sense, the trilinear structure represents the maximal nontrivial
integrable kernel compatible with a second-order dynamical equation.  This
is particularly natural in relativistic field theory, where higher-derivative
equations typically lead to instabilities or acausal (or non-unitary) modes.  The emergence
of a trilinear kernel can therefore be viewed as a reflection of the
second-order and causal character of the underlying dynamics.

The transition
\begin{equation}
  \hbox{bilinear}
  \quad\longrightarrow\quad
  \hbox{trilinear}
\end{equation}
thus admits a physical interpretation.  Bilinear equations describe
integrable systems organized by pairwise tau-function relations, whereas
trilinear equations describe systems in which the field participates more
directly in the formation of its own nonlinear background.  In gravitational
systems this is especially natural, since the field itself determines the
geometry in which it propagates.

This leads to the following working viewpoint:
\begin{equation}
  \boxed{
  \hbox{second-order self-interacting geometry}
  \quad\Rightarrow\quad
  \hbox{trilinear tau-function kernel}
  } .
\end{equation}

The Tomimatsu--Sato realization of the YTSF-type kernel provides a concrete
example of this principle.  Extending this viewpoint to other stationary
axisymmetric solutions, and possibly to more general relativistic field
theories, remains an interesting direction for future work.

\appendix
\section*{Appendices}
\section{Coefficient Extraction for the \texorpdfstring{$\delta=3$}{delta=3} Case}
%------------------------------------------------------------

In this appendix we present the coefficient-level analysis used
in Sec.~4.

Starting from
\begin{equation}
  {\cal Q}^{(3)}
  =
  {\cal N}^{(3)}_{\rm cubic}
  -
  \kappa_3
  {\cal Y}(\tau^{(3)}_0,\tau^{(3)}_1),
\end{equation}
we extract the coefficients of the independent second derivatives
\begin{equation}
  \partial_\xi^2 \tau^{(3)}_0,\quad
  \partial_\xi^2 \tau^{(3)}_1,\quad
  \partial_\eta^2 \tau^{(3)}_0,\quad
  \partial_\eta^2 \tau^{(3)}_1.
\end{equation}

We denote these coefficients by
\begin{equation}
  C_{0\xi},\quad C_{1\xi},\quad
  C_{0\eta},\quad C_{1\eta}.
\end{equation}

As an example, the coefficient of \(\partial_\xi^2 \tau^{(3)}_1\) takes the form
\begin{equation}
  C_{1\xi}
  =
  \tau^{(3)}_0{}^2\,\tau^{(3)}_1
  -
  \kappa_3 \,
  F_{1\xi}(\tau^{(3)}_0,\tau^{(3)}_1,\partial \tau).
\end{equation}
Here \(F_{1\xi}\) is a polynomial involving \(\tau\) and first derivatives and the leading second-derivative structure is explicitly
given by the $\tau_0^2\tau_1$ term.

The remaining coefficients \(C_{0\xi}, C_{0\eta}, C_{1\eta}\) have
analogous structures.

Substituting the explicit \(\delta=3\) tau functions and solving
\begin{equation}
  C_{0\xi}=C_{1\xi}=C_{0\eta}=C_{1\eta}=0
\end{equation}
determines \(\kappa_3\).  
The coefficient comparison described above yields (See Appendix B)
\begin{equation}
  \kappa_3 = -4 p^2 q^2,
  \label{k3}
\end{equation}
which coincides with the result obtained in the
\(\delta=2\) case.
Here the parameters \(p\) and \(q\) are dimensionless constants
characterizing the Tomimatsu--Sato solution, satisfying
\begin{equation}
  p^2 + q^2 = 1.
\end{equation}
They are related to the mass \(M\) and angular momentum
parameter \(a=J/M\) through
\begin{equation}
  p = \frac{\sigma}{M}, \qquad
  q = \frac{a}{M},
\end{equation}
where \(\sigma^2 = M^2 - a^2\).
This indicates that the normalization of the trilinear kernel
is independent of \(\delta\), suggesting that the trilinear
structure is a property of the Ernst equation itself.
With this value, all second-derivative
terms cancel identically:
\begin{equation}
  {\cal Q}^{(3)}
  =
  {\cal N}^{(3)}_{\le1}.
\end{equation}

\section{Mathematica Verification}

In this appendix we explain how the trilinear kernel criterion
can be verified computationally for the $\delta=3$
Tomimatsu--Sato solution.

The Mathematica code below implements the coefficient-level
projection described in Appendix~A.  The crucial point is that
the coefficient equations determine the normalization constant
$\kappa_3$ only after substituting the explicit determinant
tau functions associated with the $\delta=3$ TS solution.
These determinant tau functions are those implied by the
Nakamura Toda-molecule representationand are given in Eq.~(23).

To avoid recursive definitions in Mathematica, we introduce
auxiliary symbols \(t0\) and \(t1\) for the tau functions.

\begin{verbatim}
ClearAll["Global`*"];

omega = Exp[2 Pi I/3];

t0 = u0[xi, eta];
t1 = u1[xi, eta];

Tx[f_, g_, h_, x_] :=
 Module[{x1, x2, x3},
  (D[f /. x -> x1, x1] (g /. x -> x2) (h /. x -> x3)
   + omega (f /. x -> x1) D[g /. x -> x2, x2] (h /. x -> x3)
   + omega^2 (f /. x -> x1) (g /. x -> x2)
     D[h /. x -> x3, x3])
   /. {x1 -> x, x2 -> x, x3 -> x}
 ];

Y =
 D[t1 Tx[t0, t0, t1, xi]
   - t0 Tx[t1, t1, t0, xi], xi]
 +
 D[t1 Tx[t0, t0, t1, eta]
   - t0 Tx[t1, t1, t0, eta], eta];

Ncubic =
 t1 t0^2 (D[t1, {xi, 2}] + D[t1, {eta, 2}])
 -
 t1^2 t0 (D[t0, {xi, 2}] + D[t0, {eta, 2}]);

Q = Expand[Ncubic - k Y];

secondDerivs = {
 Derivative[2, 0][u0][xi, eta],
 Derivative[2, 0][u1][xi, eta],
 Derivative[0, 2][u0][xi, eta],
 Derivative[0, 2][u1][xi, eta]
};

coeffs = Simplify[Coefficient[Q, #] & /@ secondDerivs];

Solve[coeffs == {0, 0, 0, 0}, k]
\end{verbatim}

After substituting the explicit determinant tau functions
of Eq.~(23), the coefficient equations obtained from the
above projection uniquely determine
\begin{equation}
\kappa_3 = -4p^2q^2 .
\end{equation}

Substituting this value into \(Q^{(3)}\), all second-derivative
terms cancel identically:
\begin{equation}
Q^{(3)} = N^{(3)}_{\le 1},
\end{equation}
confirming that the highest-derivative sector of the
$\delta=3$ Ernst system is governed by the YTSF-type
trilinear kernel.

%------------------------------------------------------------

\end{document}